\documentclass[aps,pra,groupedaddress,showpacs]{revtex4}

\usepackage{graphics}
\usepackage{epsfig}
\usepackage{amsmath}

\begin{document}

\title{Intrinsic decoherence in the interaction of two fields with a two-level atom}


\author{R. Ju\'arez-Amaro}%

\affiliation{Universidad Tecnol\'ogica de la Mixteca, Apdo. Postal 71, Huajuapan de Le\'on, Oax., 69000 Mexico}
\author{J.L. Escudero-Jim\'enez and H. Moya-Cessa}
\affiliation{INAOE, Apdo. Postal 51 y 216, 72000, Puebla, Pue., Mexico}
\begin{abstract}
We study  the interaction of a two-level atom and two fields, one
of them classical. We obtain an effective Hamiltonian for this
system by using a method recently introduced that produces a small
rotation to the Hamiltonian that allows to neglect some terms in
the rotated Hamiltonian. Then we solve a variation of the
Schr\"odinger equation that models decoherence as the system
evolves through intrinsic mechanisms beyond conventional quantum
mechanics rather than dissipative interaction with an environment.
\end{abstract}
\pacs{}
\maketitle                   





\section{Introduction}
The study  the interaction between a two-level atoms and quantized
single-mode fields prepared initially in specific states such as
binomial states \cite{Orany}, displaced number states \cite{Moya},
etc. has attracted the attention over the years because of the
possibilities of engineering more interesting states of the
electromagnetic field \cite{Moya2,Moya3}. Usually the atom-field
entanglement in such states is very sensitive to dissipation and
decoherence \cite{Ge,Yu,Xu}, and the states are rapidly degraded
to statistical mixtures.

It is  complicated to find analytical solutions for problems that
include dissipation if Hamiltonians are not simplified. Such
simplification is possible under certain circumstances if the
parameters involved allow to obtain effective Hamiltonians, either
by the use of the adiabatic elimination \cite{Larson} or by the
use of other approaches such as the small rotation method
\cite{Klimov}.  When dissipation is included before effective
Hamiltonians are developed, the problem usually may only be solved
numerically. Here we apply the later method  to obtain an
effective Hamiltonian for the interaction of a two-level atom and
two fields, one quantized and the other classical and solve a
variation of the Schr\"odinger equation that models decoherence as
the system evolves through intrinsic mechanisms beyond
conventional quantum mechanics rather than dissipative interaction
with an environment \cite{Milburn}.

The atom interacting with two fields has been studied when both
fields are quantized \cite{Dutra} and for the case in which one of
them is considered classical \cite{Alsing} finding in this last
case the appearance of super-revivals \cite{Dutra2}.

\section{Atom-Two Fields Effective Hamiltonian}
The Hamiltonian for a two-level atom interacting with a quantized
field and a classical field is given by \cite{Alsing} (we set
$\hbar=1$)
\begin{equation}
H= \omega a^{\dagger}a + \omega_0 \frac{\sigma_z}{2} + \lambda
(a^{\dagger}\sigma_-+\sigma_+a ) +\epsilon e^{-i\omega t} \sigma_+
+\epsilon^* e^{i\omega t}\sigma_-
\end{equation}
where $a$ and $a^{\dagger}a$ are the annihilation and creation
operators, respectively, $\omega$ is the frequency of both
quantized and classical fields, the $\sigma$'s are the spin Pauli
matrices, $\lambda$ is the interaction constant between the atom
and the quantized field, $\omega_0$ is the atomic transition
frequency and $\epsilon$ is the amplitude of the classical field.

We may get rid off the time dependence by transforming the
Hamiltonian with $T=\exp[-i(\omega a^{\dagger}a + \omega
\frac{\sigma_z}{2} )t]$ and obtain the interaction Hamiltonian
\begin{equation}
H_{\rm I}=  \Delta \frac{\sigma_z}{2}+\lambda
(a^{\dagger}\sigma_-+\sigma_+a ) +\epsilon \sigma_+ +\epsilon^*
\sigma_-
\end{equation}
with $\Delta=\omega_0-\omega$. We consider the strong detuning
case $|\Delta|\gg \lambda, \epsilon$ and produce a {\it small
rotation} to the interaction Hamiltonian, namely, we transform it
with $R= \exp[\eta(a^{\dagger}\sigma_--\sigma_+a )]$ with the
parameter $\eta \ll 1$, such that we may approximate
$RAR^{\dagger}\approx A+ \eta[(a^{\dagger}\sigma_--\sigma_+a
),A]$, with $A$ an arbitrary operator. By taking $A=H_I$ and
setting $\eta=-\lambda/\Delta$, we obtain the effective
Hamiltonian
\begin{equation}
H_{\rm eff}=
\sigma_z\left[\frac{2\lambda^2}{\Delta}(2N+1)+\frac{2\lambda}{\Delta}(\epsilon
a^{\dagger} +\epsilon^* a)+\frac{\Delta}{2}\right]+\epsilon
\sigma_+ +\epsilon^*
 \sigma_-,\label{eff1}
\end{equation}
with $N=a^{\dagger}a$, the number operator. By displacing the
Hamiltonian above with the Glauber displacement operator
\cite{Glauber}, we obtain
\begin{equation}
H_{\rm eff}= D(\beta)\left\{\sigma_z\left[\chi N
+{\tilde{\Delta}}\right]+\epsilon \sigma_+ +\epsilon^*
 \sigma_-\right\}D^{\dagger}(\beta),\label{eff}
\end{equation}
with $\chi =-2\lambda^2/\Delta$, $\beta=\eta/\chi$  and
$\tilde{\Delta}=\Delta-|\epsilon|^2/\chi$ and its evolution
operator may be easily obtained
\begin{equation}
U_{\rm eff}(t)= D^{\dagger}(\beta)e^{-it\left(\sigma_z\left[\chi N
+{\tilde{\Delta}}\right]+\epsilon \sigma_+ +\epsilon^*
 \sigma_-\right)}D(\beta),\label{evol}
\end{equation}
with
\begin{equation}
e^{-it\left(\sigma_z\left[\frac{2\lambda^2}{\Delta}2N
+\frac{\tilde{\Delta}}{2}\right]+\epsilon \sigma_+ +\epsilon^*
 \sigma_-\right)}=
 \left[
\begin{array}{cc}
U_{\rm 11} & U_{\rm 12} \\
U_{\rm 21}& U_{\rm 22}
\end{array}
\right]
\end{equation}
where we have used the $2\times 2$ matrix representation of the
Pauli spin operators. The matrix elements are given by
\begin{eqnarray} \nonumber
 U_{\rm 11}(t) = \cos[ \Omega_Nt]-i\frac{\chi N+ \tilde{\Delta}}{\Omega_{\rm N}}\sin[ \Omega_Nt],\\
\nonumber U_{\rm 12}(t) = -i\frac{\epsilon}{\Omega_{\rm N}}\sin[ \Omega_{\rm N}t]\\
\nonumber U_{\rm 21}(t) =-i\frac{\epsilon^*}{\Omega_{\rm N}} \sin[ \Omega_{\rm N}t]\\
U_{\rm 22}(t) = \cos[ \Omega_{\rm N}t]+i\frac{\chi N+
\tilde{\Delta}}{\Omega_{\rm N}}\sin[ \Omega_{\rm N}t],
\end{eqnarray}
with $\Omega_{\rm N}=\sqrt{(\chi N+
\tilde{\Delta})^2+|\epsilon|^2}$.
\begin{figure}[hbt]
\begin{center}
\includegraphics[width=0.45\textwidth]{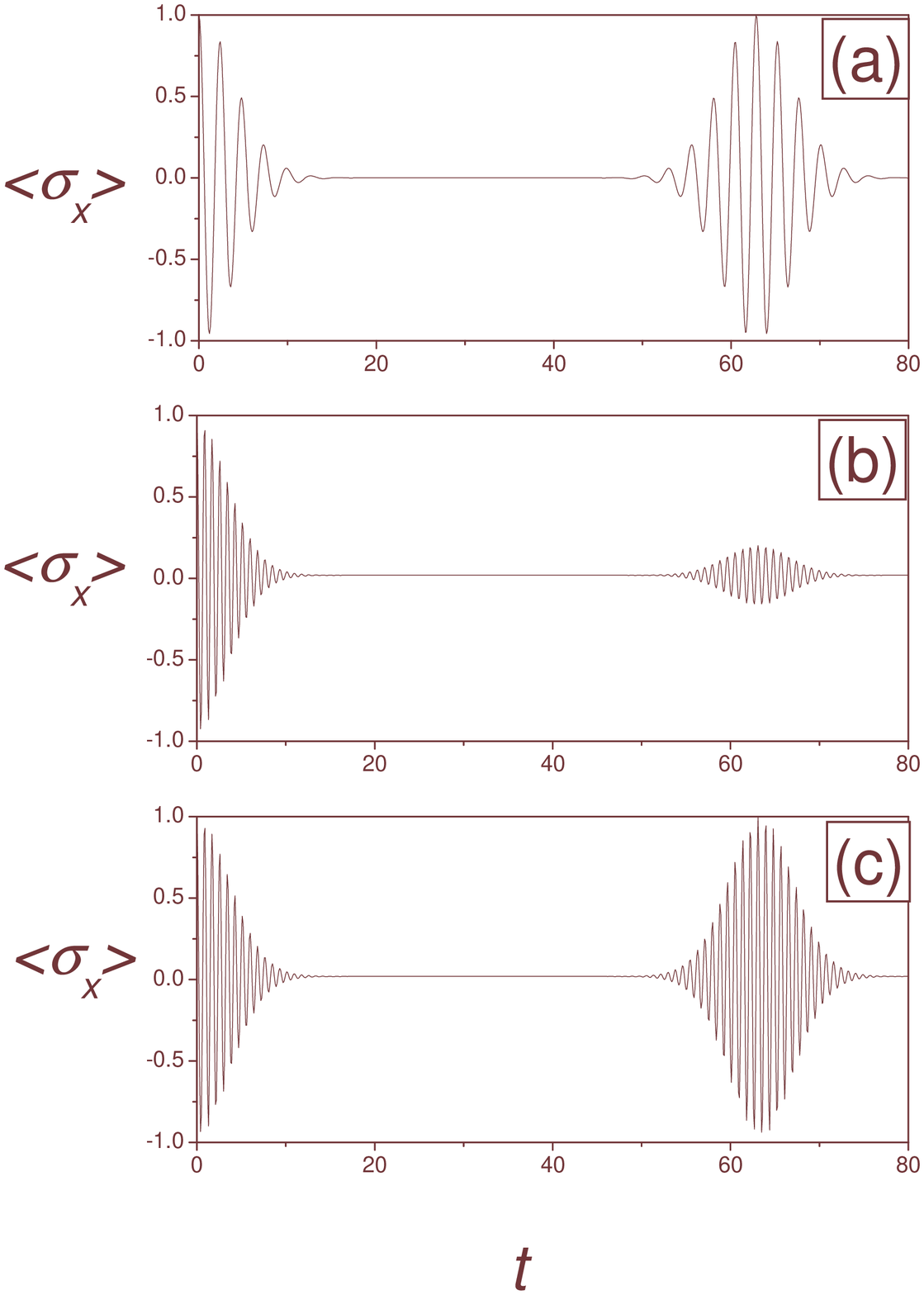}
\end{center}
\caption{\label{fig2} We plot $\langle \sigma_{\rm x} \rangle$ for
the following parameters: (a) $\alpha=2.5$, $E=0$, $\lambda=1$,
$\Delta=2$ and $\gamma=10^6$, (b)  $\alpha=2.5$, $E=0.5$,
$\lambda=1$, $\Delta=2$ and $\gamma=10^3$ and (c) $\alpha=2.5$,
$E=0.5$, $\lambda=1$, $\Delta=2$ and $\gamma=10^6$.  }
\end{figure}
\section{Intrinsic Decoherence}
There have been proposals in which the Schr\"odinger equation is
modified such that quantum coherences are destroyed as the system
evolves. Milburn \cite{Milburn} has proposed a model of intrinsic
decoherence that is a modification of quantum mechanics based on
the assumption that on sufficiently short time steps the system
does not evolve continuously under unitary evolution but rather in
an stochastic sequence of identical unitary transformations. The
differential equation for the density matrix in Milburn's model
reads
\begin{equation}
\frac{d\rho}{dt}=\gamma (e^{-iH/\hbar \gamma}\rho
e^{iH/\hbar\gamma}-\rho) \label{Milb}
\end{equation}
with $\rho$ the density matrix and $\gamma$ is the rate at which
coherences are lost and is related with the minimum time step in
the universe \cite{Milburn}. By expanding (\ref{Milb}) to first
order in $\gamma^{-1}$ Milburn obtained the following equation
\begin{equation}
\frac{d\rho}{dt}=\frac{i}{\hbar}[H,\rho]-\frac{1}{2\hbar^2\gamma}[H,[H,\rho]].
\label{Milb2}
\end{equation}
Note that taking $\gamma \rightarrow \infty$ Schr\"odinger
equation is recovered. We now solve equation (\ref{Milb}) instead
of the approximated (\ref{Milb2}) for the Hamiltonian (\ref{eff})
\begin{equation}
\rho(t)=e^{-\gamma t} \sum_{m=0}^{\infty} \frac{(\gamma t)^m}{m!}
e^{-imH_{\rm eff}/\gamma}\rho(0) e^{imH_{\rm eff}/\gamma}.
\end{equation}
By doing $t \rightarrow m/\gamma$ in  eq. (\ref{evol}), we can
easily evaluate the above solution for any initial condition for
the atom and the field. Let us consider the atom initially in a
superposition of ground and excited states
\begin{equation}
|\psi_A(0)\rangle=\frac{1}{\sqrt{2}}(|g\rangle+|e\rangle)
\end{equation}
and the field initially in a coherent state \cite{Glauber}
\begin{equation}
|\alpha\rangle=D(\alpha)|0\rangle,
\end{equation}
we may then express the initial density matrix in the atomic
$2\times 2$ basis as
\begin{equation}
\rho(0)=\frac{|\alpha\rangle\langle\alpha|}{2}
 \left[
\begin{array}{cc}
1 & 1 \\
1& 1
\end{array}
\right].
\end{equation}
We have now all the ingredients to obtain expectation values of
atomic and field observables. For instance, we can calculate the
atomic polarization $\langle \sigma_x \rangle$, a common
expectation value used in the reconstruction of the Wigner
function 
\begin{eqnarray}\nonumber
\langle \sigma_{\rm x} \rangle=&&
e^{-|\alpha-\beta|^2}\sum_{n=0}^{\infty}
\frac{|\alpha-\beta|^{2n}}{n!\Omega_{\rm n}^2}
\\&&\left(\epsilon^2+ e^{-\gamma t(1-\cos(2\Omega_{\rm
n}/\gamma))}(\chi n+\tilde{\Delta})^2\cos(\gamma
t\sin(2\Omega_{\rm n}/\gamma))\right).
\end{eqnarray}
We plot the atomic polarization in Fig. 1, where we can see
revivals and collapses of this observable (a) for the parameter
$\gamma =10^6 \lambda$, and no classical field ($\epsilon=0$);
degradation of the revivals may be seen in (b) where the parameter
$\gamma =10^3 \lambda$ has been reduced, and the classical field
is present ($\epsilon=0.5 \lambda$),  and the effects of the
intrinsic decoherence are clear and in (c) we again set $\gamma
=10^6 \lambda$ for the classical field as in (b) and the revivals
are recovered.
\section{Conclusions}
We have studied  the interaction of a two-level atom and two
fields, one of them classical in the dispersive regime by using a
model of intrinsic decoherence  that has been shown to degrade the
revivals of the atomic polarization. It was given a solution for
the exact equation \ref{Milb} rather than the approximated one
\ref{Milb2}. The dispersive Hamiltonian was obtained by using the
method of small rotations \cite{Klimov}.

  We would like to thank CONACYT for support.

\end{document}